\documentclass[a4paper,11pt]{article}
\usepackage{pos}
\usepackage{cleveref}

\newcommand{\ii}{\mathrm{i}}
\newcommand{\tr}{\operatorname{tr}}

\title{Equation of state of hot and dense QCD at the
physical point via the complex Langevin equation}
\ShortTitle{Hot and dense QCD via the complex Langevin equation}

\author[a]{Michael Mandl}
\author[a]{D{\'e}nes Sexty}
\author*[a]{Daniel Unterhuber}

\affiliation[a]{Institute of Physics, NAWI Graz, University of Graz,\\
  Universit{\"a}tsplatz 5, 8010 Graz, Austria}

\emailAdd{michael.mandl@uni-graz.at}
\emailAdd{denes.sexty@uni-graz.at}
\emailAdd{daniel.unterhuber@uni-graz.at}

\abstract{This study presents the first results for the equation of state of 2+1+1-flavor QCD at the physical point, obtained using the complex Langevin equation. We calculate continuum-extrapolated thermodynamic quantities, such as the baryon density, pressure, the speed of sound and the adiabatic compressibility. The simulations are performed in the isospin-symmetric, strangeness- and charm-neutral setting above the crossover temperature and at unprecedentedly large baryon chemical potentials.}

\FullConference{The 43rd International Symposium on Lattice Field Theory (Lattice 2026)\\
July 26 to August 1, 2026\\
University of Maryland, College Park, USA\\}

\begin{document}
\maketitle

\section{Introduction}\label{sec:introduction}
The exploration of the QCD phase diagram at finite temperature and finite baryon chemical potential is a longstanding problem in the lattice community. At $\mu_B=0$, lattice simulations have provided highly precise results, showing that, for physical quark masses, the thermal transition is not a true phase transition but a smooth analytic crossover \cite{AEF06}. The situation changes qualitatively once a finite chemical potential is introduced, as standard Monte Carlo techniques are hindered by the sign problem \cite{TW05,AS26}. A number of strategies have nevertheless been developed to access this regime, including reweighting, expansions around zero baryon chemical potential, and calculations carried out at imaginary values of $\mu_B$. Their applicability, however, deteriorates with increasing density, with current implementations typically remaining restricted to $\mu_B/T\lesssim3.5$ \cite{ADE05,BEF12,BDH17,DGS17,BDH19,GKK20,BFG21,BGK22,BFG22,BGK22_2,BFG23,BCG23,BFG24,ABF25}. A quantitative study of the high-density region of the QCD phase diagram requires approaches that circumvent the limitations of conventional importance sampling, such as the complex Langevin (CL) equation (see \cite{Par83, Kla83, Sex14, Sex14_2}). In this study, we employ the CL method to address the sign problem and perform simulations in the temperature range $260\,\mathrm{MeV}\leq T\leq600\,\mathrm{MeV}$ over a broad range of baryon chemical potentials, obtaining reliable results for ratios $\mu_B/T$ up to at least 10. While several thermodynamic quantities were already discussed in our previous work \cite{MSD26, mandl_proceeding1, mandl_proceeding2}, the present contribution additionally examines the squared speed of sound and the adiabatic compressibility.

\section{Simulation Setup}\label{sec:simulation_setup}

We use a tree-level improved Symanzik gauge action in combination with four flavors of staggered fermions (corresponding to up, down, strange, and charm quark), where the two lightest ones are considered mass degenerate. We thus effectively consider QCD with 2+1+1 flavors. In addition, we employ four steps of stout smearing with an isotropic smearing parameter of $\rho=0.125$ \cite{MP04}. To fully attack the sign problem arising at finite chemical potential, we generate gauge link configurations using the complex Langevin (CL) equation \cite{Par83, Kla83}. This approach was first applied to QCD in \cite{Sex14}. The CL method can be viewed as a stochastic evolution of complexified field variables in an additional, artificial time direction $\tau$, which is referred to as Langevin time. Upon complexifying the field manifold, the gauge links are extended from $\mathrm{SU}(3)$ to the noncompact group $\mathrm{SL(3,\mathbb{C})}$. Their stochastic evolution in $\tau$ is then realized by discretizing the corresponding continuum complex Langevin equation (see, e.g., \cite{BKK85,UF85}). An improved discretization scheme is given by
\begin{align}\label{eq:improved_langevin_step}
	\begin{aligned}
		\tilde{U}_{n, \mu}^{(\tau)} &= \exp\left[\ii\sum_a \lambda_a\left(\Delta\tau\, K_{an\mu} + \sqrt{\alpha \Delta\tau}\,\eta_{an\mu}^{(\tau)} \right)\right] U_{n, \mu}^{(\tau)}\;,\\
		U_{n, \mu}^{(\tau + \Delta\tau)} &= \exp\!\left[\ii\sum_a \lambda_a\left(\gamma\Delta\tau\, \tilde{K}_{an\mu} + \sqrt{\alpha\Delta\tau}\,\eta_{an\mu}^{(\tau)} \right)\right] U_{n, \mu}^{(\tau)}\;,
	\end{aligned}
\end{align}
where $U_{n,\mu}^{(\tau)}$ denotes the gauge link $U_{n,\mu}$ at Langevin time $\tau$, while $\Delta\tau$ specifies the corresponding discrete time step. The matrices $\lambda_a$ are the eight generators of $\mathrm{SU}(3)$, namely the Gell-Mann matrices. The stochastic contribution is represented by the real-valued Gaussian noise vector, $\eta_{an\mu}^{(\tau)}$, which has zero mean and unit variance.
The drift force follows from the Euclidean effective action of QCD and is given by
\begin{equation}\label{eq:drift_force}
    K_{an\mu} = - D_{an\mu}S_\mathrm{eff}[U] = -D_{an\mu}S_G[U] + \frac{1}{4}\sum_{f=u,d,s,c}\mathrm{tr}\left[M_f^{-1}[U]D_{an\mu} M_f[U]\right]\;,
\end{equation}
where $S_G[U]$ denotes the tree-level improved Symanzik gauge action, while $M_f[U]$ is the fermion matrix associated with flavor $f$ and depends on the corresponding quark chemical potential $\mu_f$. The left derivative of an arbitrary function $F[U]$ is defined as
\begin{equation}
    D_{an\mu}F[U] = \left.\frac{\partial F[\mathrm{e}^{\mathrm{i}\varphi \lambda_a}U_{n,\mu}]}{\partial \varphi}\right|_{\varphi=0}\;.
\end{equation}
The parameters entering the update are fixed by $\alpha = 2-3\Delta\tau$ and $\gamma=1/2 + \Delta\tau/4$, with the improved drift given by $\tilde{K}_{an\mu} = K_{an\mu}[U] + K_{an\mu}[\tilde{U}]$. An explicit calculation of the fermion determinant is not required in the CL formulation. Instead, the trace containing the inverse fermion matrix has to be evaluated (see \cref{eq:drift_force}), which we estimate stochastically, using the conjugate gradient (CG) method for the inverse. This is also the computationally most demanding part of the simulation. For a detailed description, see \cite{Sex14}. Stout smearing can likewise be extended from $\mathrm{SU}(3)$ to $\mathrm{SL}(3,\mathbb{C})$ gauge links, as described in \cite{Sex19}. For every CL simulation considered here, we furthermore apply $16$ steps of gauge cooling \cite{SSS13, NNS16_2}.

Discretizing the CL equation inevitably introduces effects associated with the finite Langevin-time step size. These would vanish only after taking the continuum limit $\Delta\tau\rightarrow0$. In practice, the improved update in \eqref{eq:improved_langevin_step} provides a sufficiently accurate approximation to this limit when $\Delta\tau$ is kept small, as confirmed by additional simulations. The step size can additionally be adjusted dynamically using an adaptive algorithm, which limits the distance traveled by a single update in configuration space \cite{AJS10}. In our simulations, we therefore combine the improved update \eqref{eq:improved_langevin_step} with an adaptive choice of $\Delta\tau$, subject to the bound $\Delta\tau \leq 0.001$, throughout the remainder of this work.

A further issue arises from zeros of the fermion determinant on the complexified manifold. Since the drift $K_{an\mu}$ contains the logarithmic derivative of the determinant, it is meromorphic, and the correctness of the CL process can no longer be taken for granted if the stochastic process comes close to the poles of the drift \cite{NS15,ASS17}. In the high-temperature regime considered in this work, however, previous studies found no indication that the Langevin trajectories approach these singularities \cite{Sex19}. The standard formal justification of the CL method \cite{ASS10,AJS11} therefore remains applicable.

Since QCD is formulated in continuous spacetime, the continuum limit is obtained on the lattice by taking $a \to 0$ and $N_s,N_t \to \infty$ while keeping the temperature, $T = 1/(aN_t)$, and the physical volume, $V = (aN_s)^3$, fixed. In this work, we use $N_t \in \{8,10,12,16\}$ and keep the aspect ratio fixed at $N_s/N_t=2$. The continuum extrapolation is performed linearly in $1/N_t^2$, excluding the $N_t=8$ data point. For simulations at fixed temperature, the bare parameters have to be tuned separately for each $N_t$ such that the corresponding lattice spacing reproduces the desired temperature and physical hadron masses. We employ the line of constant physics determined in \cite{BBF15} for 2+1+1-flavor QCD, which provides the required bare gauge coupling and quark masses for $N_t \in \{8,10,12,16\}$ at the temperatures considered here.

The infinite-volume limit requires $V\rightarrow\infty$. In this work, we do not perform an explicit volume extrapolation, but instead use a fixed aspect ratio of $N_s/N_t=2$. To check the resulting finite-volume effects, additional simulations were carried out at $N_t=8$ and $N_t=10$ with $N_s/N_t\in\{2.5,3,4\}$. The relevant physical quantities such as the baryon quark densities \eqref{eq:baryon_density} show no statistically significant dependence on the spatial volume. We therefore consider finite-volume effects to be negligible for the setup used in this work.

Observables are measured every $0.1$ units of Langevin time in all simulations. For almost all parameter sets $\{N_t,T,\mu_l\}$, two independent runs with different random number seeds are performed with a total number of generated samples in the order of $\mathcal{O}(10^3)$. Here, $\mu_l$ denotes the light-quark chemical potential in lattice units; further details are given in \cref{sec:results}. For the baryon density $n_B$ (see \eqref{eq:baryon_density}), thermal equilibrium is reached after approximately $6$ units of Langevin time for all temperatures and temporal extents considered, with somewhat longer equilibration times at larger $\mu_l$. We therefore introduce the minimal thermalization cut of $\delta\tau=7$, by discarding the data of the first $7$ units of Langevin in the subsequent analysis. Statistical uncertainties are estimated using jackknife resampling with blocking. The block size is chosen to be at least six times the integrated autocorrelation time, ensuring that residual autocorrelations are negligible. Further details of the error analysis are given in \cref{sec:results}.

\section{Results}\label{sec:results}

As stated in the introduction (see \cref{sec:introduction}), the simulations are performed in the isospin-symmetric, strangeness- and charm-neutral case, corresponding to the conditions
\begin{equation}
    \mu_u = \mu_d \equiv \mu_l, \quad \mu_s = \mu_c = 0\;.
\end{equation}
It should be noted that this choice does not exactly correspond to the strangeness- and charm-neutral condition at low temperatures. However, for the temperatures considered in this study, $T \geq 260\;\mathrm{MeV}$, the difference is negligible to a very good approximation (see \cite{BGK22}).

\begin{figure}[h!]
    \centering
    \begin{minipage}[t]{0.49\linewidth}
        \centering
        \includegraphics[width=\linewidth]{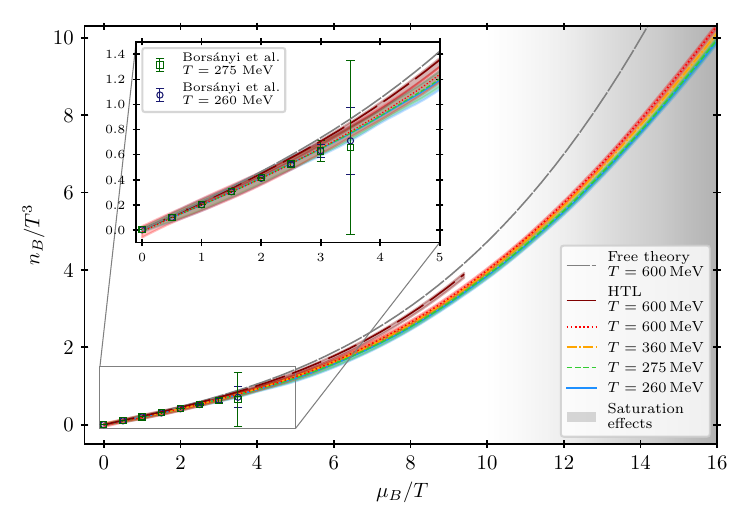}
    \end{minipage}
    \hfill
    \begin{minipage}[t]{0.49\linewidth}
        \centering
        \includegraphics[width=\linewidth]{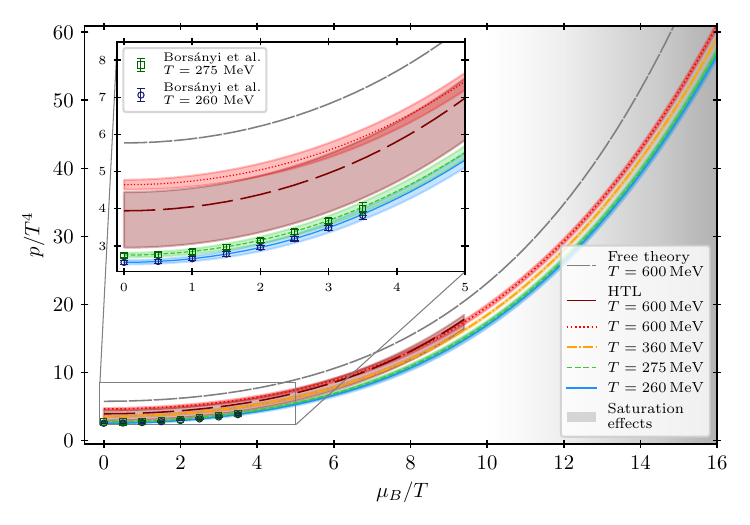}
    \end{minipage}
    \caption{Continuum-extrapolated baryon density, $n_B$ (left), and pressure, $p$ (right), as functions of the baryon chemical potential, $\mu_B$, for several temperatures, $T$. Both quantities are normalized by the corresponding powers of $T$. In the small-$\mu_B$ regime, the results are compared with other lattice calculations \cite{BGK22} and with hard-thermal-loop (HTL) perturbation theory. The gray dashed curves indicate the corresponding free-theory predictions for quarks and gluons. The insets show enlarged views of the small-$\mu_B$ region. The shaded areas indicate the region in which saturation effects become relevant; further details are provided in the main text. Figure taken from~\cite{MSD26}.}
    \label{fig:baryon_density_pressure_baryon_mu}
\end{figure}

The baryon density, $n_B(T,\mu_B)$, as a function of the temperature, $T$, and the baryon chemical potential, $\mu_B = 3\mu_l$, can be measured directly from the lattice configurations according to
\begin{equation}\label{eq:baryon_density}
    n_B(T,\mu_B) = \frac{T}{V}\frac{\partial\ln\mathcal{Z}}{\partial\mu_B}
    = \frac{T}{12V}\sum_{f=u,d,s,c}
    \left\langle\tr\left(M_f^{-1}\frac{\partial M_f}{\partial\mu_f}\right)\right\rangle\;,
\end{equation}
where $\tr(\cdot)$ denotes the trace and $\mathcal{Z}$ is the partition function. Following the continuum extrapolation and interpolation in the $T$- and $\mu_B$-directions, the resulting baryon density, normalized by $T^3$ to yield a dimensionless quantity, is shown on the left-hand side of \cref{fig:baryon_density_pressure_baryon_mu} for several fixed temperatures $T$. It increases with increasing baryon chemical potential, whereas the temperature dependence of $n_B/T^3$ is very mild. In \cref{fig:baryon_density_pressure_baryon_mu}, our results are compared with other available lattice results~\cite{BGK22} and with perturbative calculations based on hard-thermal-loop (HTL) theory~\cite{HAM14,HBA14,AHM16}. For further comparison, the corresponding free-theory result, consisting of non-interacting Bose and Fermi gases, is also displayed. The figure demonstrates that the CL approach allows us to obtain non-perturbative results at substantially larger baryon chemical potentials than are typically accessible with conventional importance-sampling methods such as hybrid Monte Carlo (HMC).

The uncertainty bands of our results are obtained as follows. To quantify the systematic uncertainty, the computation of the baryon density is repeated while varying four independent analysis choices: the thermalization threshold, $\delta\tau \in \{7,8,9,10\}$; the interpolation scheme in the $\mu_B$-direction, using either a cubic spline or a ninth-degree polynomial containing only odd powers; the subset of data points included in the interpolation, consisting of either all, even, or odd points; and the number of jackknife blocks, $n_\mathrm{blocks} \in \{10,12,14,16\}$. The combination of these variations yields a total of $96$ different analysis results. The median of these $96$ determinations is used as the central value. To obtain a conservative estimate of the statistical uncertainty, we take the largest statistical error among all $96$ analyses. The systematic uncertainty is estimated as half the width of the central $68\%$ interval of the resulting distribution. The final uncertainty is then obtained by adding the statistical and systematic uncertainties linearly.

For sufficiently large baryon chemical potentials, the results become increasingly unreliable, as indicated by the gray shaded region in \cref{fig:baryon_density_pressure_baryon_mu}. This limitation is related to the continuum extrapolation: the $\chi_\mathrm{red}^2$ of the linear fit becomes large, indicating a poor fit quality and, consequently, an unreliable extrapolation. The underlying cause is the onset of saturation effects on the lattices with the smallest number of lattice sites. These effects arise from the finite number of fermionic modes available on a discretized lattice. At sufficiently large chemical potentials, the available modes become fully occupied, so that the baryon density eventually ceases to increase.

The pressure difference can be obtained from the baryon density via
\begin{equation}\label{eq:pressure_difference}
    \Delta p(T,\mu_B)=\int_0^{\mu_B} d\mu_B'\;n_B(T,\mu_B')\;,
\end{equation}
and the total pressure is then given by
\begin{equation}
    p(T,\mu_B)=p(T,0) +\Delta p(T,\mu_B)\;.
\end{equation}
The integral in \cref{eq:pressure_difference} is evaluated numerically, while the pressure at zero chemical potential, $p(T,0)$, is taken from the supplementary material of \cite{BFG16}. The resulting total pressure, normalized by $T^4$, is shown on the right-hand side of \cref{fig:baryon_density_pressure_baryon_mu}. The free-theory prediction (gray dashed line) lies above all other results throughout the considered range, indicating that the QCD medium cannot be described by a free theory, even at such high temperatures and baryon chemical potentials.

Using additional thermodynamic relations, one can derive further quantities, such as the trace anomaly, $I$, the energy density, $\varepsilon$, the entropy density, $s$, and the enthalpy density, $w$ (see, e.g., \cite{mandl_proceeding1, mandl_proceeding2}). In this study, the squared speed of sound is calculated using
\begin{equation}\label{eq:speed_of_sound}
    c_s^2 = \left.\frac{\partial p}{\partial \varepsilon}\right|_{s/n_B=\mathrm{const.}} 
    = \left. \frac{T \left( \frac{\partial (p/T^4)}{\partial T} + \frac{\partial (p/T^4)}{\partial (\mu_B/T)} \frac{\mathrm{d} (\mu_B/T)}{\mathrm{d} T} \right) + 4 (p/T^4)}
    {T \left( \frac{\partial (\varepsilon/T^4)}{\partial T} + \frac{\partial (\varepsilon/T^4)}{\partial (\mu_B/T)} \frac{\mathrm{d} (\mu_B/T)}{\mathrm{d} T} \right) + 4 (\varepsilon/T^4)} \right|_{s/n_B=\mathrm{const.}}\;,
\end{equation}
where all quantities are evaluated along trajectories of constant entropy per baryon density, $s/n_B$.

\begin{figure}[h!]
    \centering
    \begin{minipage}[t]{0.49\linewidth}
        \centering
        \includegraphics[width=\linewidth]{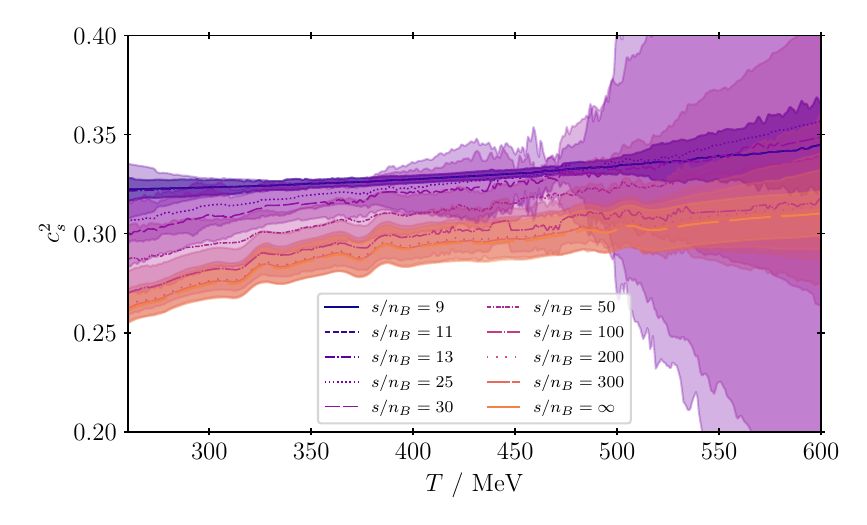}
    \end{minipage}
    \hfill
    \begin{minipage}[t]{0.49\linewidth}
        \centering
        \includegraphics[width=\linewidth]{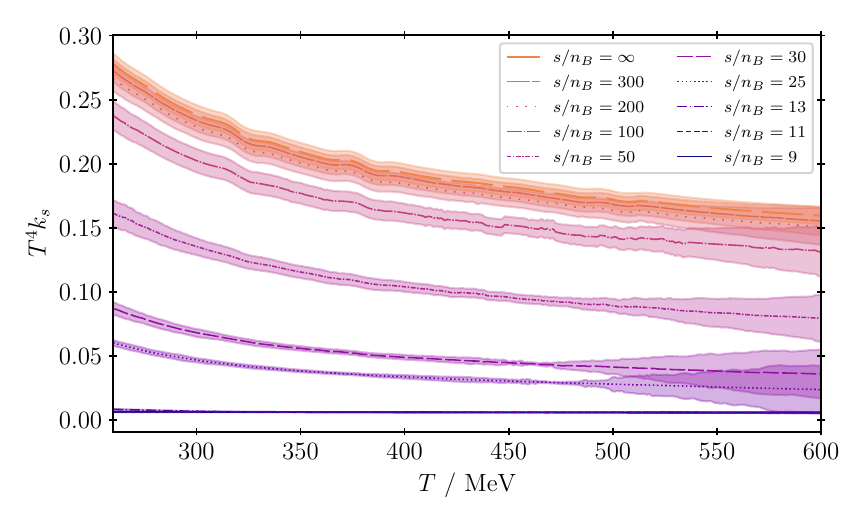}
    \end{minipage}
    \caption{Continuum-extrapolated squared speed of sound, $c_s^2$ (left), and adiabatic compressibility, $T^4 k_s$ (right), as functions of the temperature, $T$, for several fixed values of $s/n_B$.}
    \label{fig:speed_of_sound_adiabatic_compressibility}
\end{figure}

Finally, the adiabatic compressibility is calculated as
\begin{equation}\label{eq:adiabatic_compressibility}
    T^4 k_s = \left.\frac{1}{c_s^2\,(w/T^4)}\right|_{s/n_B=\mathrm{const.}}\;.
\end{equation}
The results for both quantities are shown in \cref{fig:speed_of_sound_adiabatic_compressibility}. Note that $s/n_B = \infty$ corresponds to zero chemical potential, while decreasing values of $s/n_B$ correspond to increasing baryon chemical potentials. Within the considered temperature range, both quantities exhibit the qualitatively expected behavior (see \cite{BCG23} for a comparison extending up to $T=300\textrm{ MeV}$). Note that the observed wiggles are numerical artifacts arising from the differentiation and/or interpolation of noisy data.

\section{Conclusion}

This work presents the first results for QCD at the physical point obtained using the complex Langevin equation. We have calculated the equation of state of 2+1+1-flavor QCD at high temperatures ($T\geq260$ MeV) and baryon chemical potentials up to $\mu_B/T\lesssim10$. These continuum-extrapolated results represent the first non-perturbative computations at such large chemical potentials. In particular we computed the baryon density, the pressure, the squared speed of sound and the adiabatic compressibility.

\section*{Acknowledgments}

We thank Astrid Carstensen, Erhard Seiler, and Ion-Olimpiu Stamatescu for valuable discussions on this and related topics. We are also grateful to Szabolcs Borsányi and Jana Günther for providing the data at vanishing chemical potential and the data for the line of constant physics. D.U. particularly thanks Jana Günther for fruitful discussions at Lattice 2026. The computational results have been achieved using the Austrian Scientific Computing (ASC) infrastructure as well as the computing cluster of the university of Graz (GSC). This research was funded in whole or in part by the Austrian Science Fund (FWF) \href{https://doi.org/10.55776/P36875}{10.55776/P36875}.

\bibliographystyle{JHEP}
\bibliography{bibliography}

@article{AEF06,
	author = "Aoki, Y. and Endr\H{o}di, G. and Fodor, Z. and Katz, S. D. and Szab{\'o}, K. K.",
	title = "{The order of the quantum chromodynamics transition predicted by the standard model of particle physics}",
	journal = "Nature",
	volume = "443",
	pages = "675--678",
	doi = "10.1038/nature05120",
	eprint = "hep-lat/0611014",
	archivePrefix = "arXiv",
	year = "2006"
}

@article{TW05,
	author = "Troyer, M. and Wiese, U.-J.",
	title = "{Computational Complexity and Fundamental Limitations to Fermionic Quantum Monte Carlo Simulations}",
	journal = "Phys. Rev. Lett.",
	volume = "94",
	pages = "170201",
	doi = "10.1103/PhysRevLett.94.170201",
	eprint = "cond-mat/0408370",
	archivePrefix = "arXiv",
	year = "2005"
}

@article{AS26,
    author = "Aarts, Gert and Sexty, D{\'e}nes",
    title = "{Lattice field theories with a sign problem}",
    eprint = "2604.24290",
    archivePrefix = "arXiv",
    primaryClass = "hep-lat",
    month = "4",
    year = "2026"
}

@article{ADE05,
	author = "Allton, C. R. and D{\"o}ring, M. and Ejiri, S. and Hands, S. J. and Kaczmarek, O. and Karsch, F. and Laermann, E. and Redlich, K.",
	title = "{Thermodynamics of two flavor QCD to sixth order in quark chemical potential}",
	journal = "Phys. Rev. D",
	volume = "71",
	pages = "054508",
	doi = "10.1103/PhysRevD.71.054508",
	eprint = "hep-lat/0501030",
	archivePrefix = "arXiv",
	year = "2005"
}

@article{BEF12,
	author = "Bors{\'a}nyi, Sz. and Endr\H{o}di, G. and Fodor, Z. and Katz, S. D. and Krieg, S. and Ratti, C. and Szab{\'o}, K. K.",
	title = "{QCD equation of state at nonzero chemical potential: continuum results with physical quark masses at order $\mu^2$}",
	journal = "JHEP",
	volume = "2012",
	number = "053",
	doi = "10.1007/JHEP08(2012)053",
	eprint = "1204.6710",
	archivePrefix = "arXiv",
	primaryClass = "hep-lat",
	year = "2012"
}

@article{BDH17,
	author = "Bazavov, A. and Ding, H.-T. and Hegde, P. and Kaczmarek, O. and Karsch, F. and Laermann, E. and Maezawa, Y. and Mukherjee, S. and Ohno, H. and Petreczky, P. and Sandmeyer, H. and Steinbrecher, P. and Schmidt, C. and Sharma, S. and Soeldner, W. and Wagner, M.",
	title = "{QCD equation of state to $\mathcal{O}(\mu_B^6)$ from lattice QCD}",
	journal = "Phys. Rev. D",
	volume = "95",
	pages = "054504",
	doi = "10.1103/PhysRevD.95.054504",
	eprint = "1701.04325",
	archivePrefix = "arXiv",
	primaryClass = "hep-lat",
	year = "2017"
}

@article{DGS17,
	author = "D'Elia, M. and Gagliardi, G. and Sanfilippo, F.",
	title = "{Higher order quark number fluctuations via imaginary chemical potentials in $N_f=2+1$ QCD}",
	journal = "Phys. Rev. D",
	volume = "95",
	pages = "094503",
	doi = "10.1103/PhysRevD.95.094503",
	eprint = "1611.08285",
	archivePrefix = "arXiv",
	primaryClass = "hep-lat",
	year = "2017"
}

@article{BDH19,
	author = "\relax{A. Bazavov et al. (HotQCD Collaboration)}",
	title = "{Chiral crossover in QCD at zero and non-zero chemical potentials}",
	journal = "Phys. Lett. B",
	volume = "795",
	pages = "15--21",
	doi = "10.1016/j.physletb.2019.05.013",
	eprint = "1812.08235",
	archivePrefix = "arXiv",
	primaryClass = "hep-lat",
	year = "2019"
}

@article{GKK20,
	author = "Giordano, M. and Kapas, K. and Katz, S. D. and Nogradi, D. and Pasztor, A.",
	title = "{New approach to lattice QCD at finite density; results for the critical end point on coarse lattices}",
	journal = "JHEP",
	volume = "2020",
	number = "088",
	doi = "10.1007/JHEP05(2020)088",
	eprint = "2004.10800",
	archivePrefix = "arXiv",
	primaryClass = "hep-lat",
	year = "2020"
}

@article{BFG21,
	author = "Bors{\'a}nyi, Sz. and Fodor, Z. and Guenther, J. N. and Kara, R. and Katz, S. D. and Parotto, P. and P{\'a}sztor, A. and Ratti, C. and Szab{\'o}, K. K.",
	title = "{Lattice QCD Equation of State at Finite Chemical Potential from an Alternative Expansion Scheme}",
	journal = "Phys. Rev. Lett.",
	volume = "126",
	pages = "232001",
	doi = "10.1103/PhysRevLett.126.232001",
	eprint = "2102.06660",
	archivePrefix = "arXiv",
	primaryClass = "hep-lat",
	year = "2021"
}

@article{BGK22,
	author = "Bors{\'a}nyi, Sz. and Guenther, J. N. and Kara, R. and Fodor, Z. and Parotto, P. and P{\'a}sztor, A. and Ratti, C. and Szab{\'o}, K.",
	title = "{Resummed lattice QCD equation of state at finite baryon density: Strangeness neutrality and beyond}",
	journal = "Phys. Rev. D",
	volume = "105",
	pages = "114504",
	doi = "10.1103/PhysRevD.105.114504",
	eprint = "2202.05574",
	archivePrefix = "arXiv",
	primaryClass = "hep-lat",
	year = "2022"
}

@article{HAM14,
	author = "Haque, N. and Andersen, J. O. and Mustafa, M. G. and Strickland, M. and Su, N.",
	title = "{Three-loop pressure and susceptibility at finite temperature and density from hard-thermal-loop perturbation theory}",
	journal = "Phys. Rev. D",
	volume = "89",
	pages = "061701(R)",
	doi = "10.1103/PhysRevD.89.061701",
	eprint = "1309.3968",
	archivePrefix = "arXiv",
	primaryClass = "hep-ph",
	year = "2014"
}

@article{HBA14,
	author = "Haque, N. and  Bandyopadhyay, A. and Andersen, J. O. and Mustafa, M. G. and Strickland, M. and Su, N.",
	title = "{Three-loop HTLpt thermodynamics at finite temperature and chemical potential}",
	journal = "JHEP",
	volume = "2014",
	number = "027",
	doi = "10.1007/JHEP05(2014)027",
	eprint = "1402.6907",
	archivePrefix = "arXiv",
	primaryClass = "hep-ph",
	year = "2014"
}

@article{AHM16,
	author = "Andersen, J. O. and Haque, N. and Mustafa, M. G. and Strickland, M.",
	title = "{Three-loop hard-thermal-loop perturbation theory thermodynamics at finite temperature and finite baryonic and isospin chemical potential}",
	journal = "Phys. Rev. D",
	volume = "93",
	pages = "054045",
	doi = "10.1103/PhysRevD.93.054045",
	eprint = "1511.04660",
	archivePrefix = "arXiv",
	primaryClass = "hep-ph",
	year = "2016"
}

@article{BFG22,
	author = "Bors{\'a}nyi, Sz. and Fodor, Z. and Giordano, M. and Katz, S. D. and N{\'o}gradi, D. and P{\'a}sztor, A. and Wong, C. H.",
	title = "{Lattice simulations of the QCD chiral transition at real baryon density}",
	journal = "Phys. Rev. D",
	volume = "105",
	pages = "L051506",
	doi = "10.1103/PhysRevD.105.L051506",
	eprint = "2108.09213",
	archivePrefix = "arXiv",
	primaryClass = "hep-lat",
	year = "2022"
}

@article{BGK22_2,
	author = "\relax{D. Bollweg et al. (HotQCD Collaboration)}",
	title = "{Taylor expansions and Pad{\'e} approximants for cumulants of conserved charge fluctuations at nonvanishing chemical potentials}",
	journal = "Phys. Rev. D",
	volume = "105",
	pages = "074511",
	doi = "10.1103/PhysRevD.105.074511",
	eprint = "2202.09184",
	archivePrefix = "arXiv",
	primaryClass = "hep-lat",
	year = "2022"
}

@article{BFG23,
	author = "Bors{\'a}nyi, Sz. and Fodor, Z. and Giordano, M. and Guenther, J. N. and Katz, S. D. and P{\'a}sztor, A. and Wong, C. H.",
	title = "{Equation of state of a hot-and-dense quark gluon plasma: Lattice simulations at real $\mu_B$ vs extrapolations}",
	journal = "Phys. Rev. D",
	volume = "107",
	pages = "L091503",
	doi = "10.1103/PhysRevD.107.L091503",
	eprint = "2208.05398",
	archivePrefix = "arXiv",
	primaryClass = "hep-lat",
	year = "2023"
}

@article{BCG23,
	author = "\relax{D. Bollweg et al. (HotQCD Collaboration)}",
	title = "{Equation of state and speed of sound of (2+1)-flavor QCD in strangeness-neutral matter at nonvanishing net baryon-number density}",
	journal = "Phys. Rev. D",
	volume = "108",
	pages = "014510",
	doi = "10.1103/PhysRevD.108.014510",
	eprint = "2212.09043",
	archivePrefix = "arXiv",
	primaryClass = "hep-lat",
	year = "2023"
}

@article{BFG24,
	author = "Bors{\'a}nyi, Sz. and Fodor, Z. and Guenther, J. N. and Parotto, P. and P{\'a}sztor, A. and Pirelli, L. and Szab{\'o}, K. K. and Wong, C. H.",
	title = "{QCD deconfinement transition line up to $\mu_B=400 MeV$ from finite volume lattice simulations}",
	journal = "Phys. Rev. D",
	volume = "110",
	pages = "114507",
	doi = "10.1103/PhysRevD.110.114507",
	eprint = "2410.06216",
	archivePrefix = "arXiv",
	primaryClass = "hep-lat",
	year = "2024"
}

@article{ABF25,
	author = "Abuali, A. and Bors{\'a}nyi, Sz. and Fodor, Z. and Jahan, J. and Kahangirwe, M. and Parotto, P. and P{\'a}sztor, A. and Ratti, C. and Shah, H. and Trabulsi, S. A.",
	title = "{New 4D lattice QCD equation of state: Extended density coverage from a generalized $T'$ expansion}",
	journal = "Phys. Rev. D",
	volume = "112",
	pages = "054502",
	doi = "10.1103/2dmh-26yh",
	eprint = "2504.01881",
	archivePrefix = "arXiv",
	primaryClass = "hep-lat",
	year = "2025"
}

@article{Par83,
	author = "Parisi, G.",
	title = "{On complex probabilities}",
	journal = "Phys. Lett. B",
	volume = "131",
	pages = "393--395",
	doi = "10.1016/0370-2693(83)90525-7",
	year = "1983"
}

@article{Kla83,
	author = "Klauder, J. R.",
	title = "{A Langevin approach to fermion and quantum spin correlation functions}",
	journal = "J. Phys. A",
	volume = "16",
	pages = "L317",
	doi = "10.1088/0305-4470/16/10/001",
	year = "1983"
}

@article{SSS13,
	author = "Seiler, E. and Sexty, D. and Stamatescu, I.-O.",
	title = "{Gauge cooling in complex Langevin for lattice QCD with heavy quarks}",
	journal = "Phys. Lett. B",
	volume = "723",
	pages = "213--216",
	doi = "10.1016/j.physletb.2013.04.062",
	eprint = "1211.3709",
	archivePrefix = "arXiv",
	primaryClass = "hep-lat",
	year = "2013"
}

@article{NNS16_2,
	author = "Nagata, K. and Nishimura, J. and Shimasaki, S.",
	title = "{Justification of the complex Langevin method with the gauge cooling procedure}",
	journal = "Prog. Theor. Exp. Phys.",
	volume = "2016",
	pages = "013B01",
	doi = "10.1093/ptep/ptv173",
	eprint = "1508.02377",
	archivePrefix = "arXiv",
	primaryClass = "hep-lat",
	year = "2016"
}

@article{Sex14,
	author = "Sexty, D.",
	title = "{Simulating full QCD at nonzero density using the complex Langevin equation}",
	journal = "Phys. Lett B.",
	volume = "729",
	pages = "108--111",
	doi = "10.1016/j.physletb.2014.01.019",
	eprint = "1307.7748",
	archivePrefix = "arXiv",
	primaryClass = "hep-lat",
	year = "2014"
}

@article{Sex14_2,
	author = "Sexty, D.",
	title = "{Progress in complex Langevin simulations of full QCD at non-zero density}",
	journal = "Nucl. Phys. A",
	volume = "931",
	pages = "856--860",
	doi = "10.1016/j.nuclphysa.2014.09.029",
	eprint = "1408.6767",
	archivePrefix = "arXiv",
	primaryClass = "hep-lat",
	year = "2014"
}

@article{ASS17,
	author = "Aarts, G. and Seiler, E. and Sexty, D. and Stamatestu, I.-O.",
	title = "{Complex Langevin dynamics and zeroes of the fermion determinant}",
	journal = "JHEP",
	volume = "2017",
	number = "044",
	doi = "10.1007/JHEP05(2017)044",
	eprint = "1701.02322",
	archivePrefix = "arXiv",
	primaryClass = "hep-lat",
	year = "2017",
}

@article{Sex19,
	author = "Sexty, D.",
	title = "{Calculating the equation of state of dense quark-gluon plasma using the complex Langevin equation}",
	journal = "Phys. Rev. D",
	volume = "100",
	pages = "074503",
	doi = "10.1103/PhysRevD.100.074503",
	eprint = "1907.08712",
	archivePrefix = "arXiv",
	primaryClass = "hep-lat",
	year = "2019",
}

@article{ASS10,
	author = "Aarts, G. and Seiler, E. and Stamatescu, I.-O.",
	title = "{Complex Langevin method: When can it be trusted?}",
	journal = "Phys. Rev. D",
	volume = "81",
	pages = "054508",
	doi = "10.1103/PhysRevD.81.054508",
	eprint = "0912.3360",
	archivePrefix = "arXiv",
	primaryClass = "hep-lat",
	year = "2010"
}

@article{AJS11,
	author = "Aarts, G. and James, F. A. and Seiler, E. and Stamatescu, I.-O.",
	title = "{Complex Langevin: etiology and diagnostics of its main problem}",
	journal = "Eur. Phys. J. C",
	volume = "71",
	pages = "1756",
	doi = "10.1140/epjc/s10052-011-1756-5",
	eprint = "1101.3270",
	archivePrefix = "arXiv",
	primaryClass = "hep-lat",
	year = "2011"
}

@article{MSD26,
    author = "Mandl, Michael and Sexty, D{\'e}nes and Unterhuber, Daniel",
    title = "{Finite-density equation of state of hot QCD using the complex Langevin equation}",
    eprint = "2604.19649",
    archivePrefix = "arXiv",
    primaryClass = "hep-lat",
    month = "4",
    year = "2026"
}

@article{AJS10,
	author = "Aarts, G. and James, F. A. and Seiler, E. and Stamatescu, I.-O.",
	title = "{Adaptive stepsize and instabilities in complex Langevin dynamics}",
	journal = "Phys. Lett. B",
	volume = "687",
	pages = "154--159",
	doi = "10.1016/j.physletb.2010.03.012",
	eprint = "0912.0617",
	archivePrefix = "arXiv",
	primaryClass = "hep-lat",
	year = "2010"
}

@article{BKK85,
	author = "Batrouni, G. G. and Katz, G. R. and Kronfeld, A. S. and Lepage, G. P. and Svetitsky, B. and Wilson, K. G.",
	title = "{Langevin simulations of lattice field theories}",
	journal = "Phys. Rev. D",
	volume = "32",
	pages = "2736",
	doi = "10.1103/PhysRevD.32.2736",
	year = "1985"
}

@article{UF85,
	author = "Ukawa, A. and Fukugita, M.",
	title = "{Langevin Simulation Including Dynamical Quark Loops}",
	journal = "Phys. Rev. Lett.",
	volume = "55",
	pages = "1854",
	doi = "10.1103/PhysRevLett.55.1854",
	year = "1985"
}

@article{MP04,
	author = "Morningstar, C. and Peardon, M.",
	title = "{Analytic smearing of $SU(3)$ link variables in lattice QCD}",
	journal = "Phys. Rev. D",
	volume = "69",
	pages = "054501",
	doi = "10.1103/PhysRevD.69.054501",
	eprint = "hep-lat/0311018",
	archivePrefix = "arXiv",
	year = "2004"
}

@article{BBF15,
	author = "Bellwied, R. and Bors{\'a}nyi, S. and Fodor, Z. and Katz, S. D. and P{\'a}sztor, A. and Ratti, C. and Szab{\'o}, K. K.",
	title = "{Fluctuations and correlations in high temperature QCD}",
	journal = "Phys. Rev. D",
	volume = "92",
	pages = "114505",
	doi = "10.1103/PhysRevD.92.114505",
	eprint = "1507.04627",
	archivePrefix = "arXiv",
	primaryClass = "hep-lat",
	year = "2015"
}

@article{BFG16,
	author = "Borsanyi, S. and Fodor, Z. and Guenther, J. and Kampert, K.-H. and Katz, S. D. and Kawanai, T. and Kovacs, T. G. and Mages, S. W. and Pasztor, A. and Pittler, F. and Redondo, J. and Ringwald, A. and Szabo, K. K.",
	title = "{Calculation of the axion mass based on high-temperature quantum chromodynamics}",
	journal = "Nature",
	volume = "539",
	pages = "69--71",
	doi = "10.1038/nature20115",
	eprint = "1606.07494",
	archivePrefix = "arXiv",
	primaryClass = "hep-lat",
	year = "2016"
}

@article{NS15,
	author = "Nishimura, J. and Shimasaki, S.",
	title = "{New insights into the problem with a singular drift term in the complex Langevin method}",
	journal = "Phys. Rev. D",
	volume = "92",
	pages = "011501(R)",
	doi = "10.1103/PhysRevD.92.011501",
	eprint = "1504.08359",
	archivePrefix = "arXiv",
	primaryClass = "hep-lat",
	year = "2015"
}

@article{mandl_proceeding1,
      title="Equation of state of hot and dense QCD using the complex Langevin equation", 
      author="Mandl, Michael and Sexty, D{\'e}nes and Unterhuber, Daniel",
      year="2026",
      eprint="2610.32385",
      archivePrefix="arXiv",
      primaryClass="hep-lat",
      url="https://arxiv.org/abs/2610.32385", 
}

@article{mandl_proceeding2,
      title="Hot and dense QCD equation of state from the complex Langevin equation", 
      author="Mandl, Michael and Sexty, D{\'e}nes and Unterhuber, Daniel",
      year="2026",
      eprint="2610.38466",
      archivePrefix="arXiv",
      primaryClass="hep-lat",
      url="https://arxiv.org/abs/2610.38466", 
}

%\begin{thebibliography}{99}
%\bibitem{...}
%....
%
%\end{thebibliography}

\end{document}